\documentclass{emulateapj}
\usepackage{apjfonts}
\lefthead{PARK \& HAN}
\righthead{MICROLENING COLOR SHIFT}

\begin{document}
\title{Color-Shift Measurement in Microlensing-Induced Stellar 
Variation from Future Space-Based Surveys}

\author{Byeong-Gon Park}
\affil{
Korea Astronomy and Space Science Institute, 838 Daedeok-Rho, 
Yuseong-Gu, Daejeon 305-348, Korea;\\
bgpark@kasi.re.kr}

\and

\author{Cheongho Han\footnote{corresponding author}}
\affil{Program of Brain Korea 21, Astrophysical Research Center 
for the Structure and Evolution of the Cosmos, SRC, \\Department of Physics, 
Chungbuk National University, Chongju 361-763, Korea;\\
cheongho@astroph.chungbuk.ac.kr}

\submitted{Submitted to The Astrophysical Journal}

\begin{abstract}

If a microlensing event is caused by a star, the event can 
exhibit change in color due to the light from the lens.  In 
the previous and current lensing surveys, the color shift
could not be used to constrain the lens population because 
the blended light responsible for the color shift is mostly 
attributed to nearby background stars rather than the lens.
However, events to be observed in future space-based surveys  
do not suffer from blending and thus the color information can 
be used to constrain lenses.  In this paper, we demonstrate 
the usefulness of future surveys in measuring color shifts.  
By conducting simulation of galactic lensing events based on 
the specification of a proposed space-based lensing  survey, 
we estimate that the shift in the color of $R-H$ will be 
measured at 5$\sigma$ level for $\sim 12\%$ of events that
occur on source stars with apparent magnitudes brighter 
than $J=22.5$.  Color-shifted events tend to have high
magnifications and the lenses will have brightnesses equivalent 
to those of source stars.  The time scales of the color-shifted 
events tend to be longer than those without color shifts. From 
the mass distribution of lenses, we find that most of the 
color-shifted events will be produced by stellar lenses with 
spectral types down to mid M-type main sequence stars. 
\end{abstract}

\keywords{gravitational lensing}


\section{Introduction}

One important characteristic of a microlensing event is that 
the color of the lensed star does not change during the lensing 
magnification.  However, events observed in actual lensing 
surveys can exhibit color shift.  This color shift occurs when 
the flux from the lensed source star is blended with the flux 
from other stars located in the seeing disk of the star.  The 
lensed and blended stars generally have different colors.
Then, color shift occurs as the source/blend relative brightness 
changes during lensing magnification.

\citet{kamionkowski95} and \citet{buchalter96} pointed out that 
if the object responsible for a lensing event is a star, it is 
possible to obtain information about the lens from the measured 
color shift.  However, this idea could not be implemented for 
the actual analysis of lens population.  The most important reason 
for this is that under the previous and current lensing surveys 
conducted by using ground-based telescopes, the blended flux is 
mostly attributed to nearby background stars rather than the lens.

However, the situation will be different in future space-based 
lensing surveys, such as the {\it Microlensing Planet Finder} 
({\it MPF}).  The {\it MPF} is a proposed space mission to NASA's 
Discovery Program exclusive for microlensing and it will be 
equipped with a 1.1 m aperture telescope \citep{bennett02, bennett04}.  
Such a survey is ideal for the measurement of the color shift in 
lensing events due to the following reasons.  First, thanks to 
the high resolution provided from space observation, the survey 
does not suffer from blending caused by background stars.  Then, 
one can safely attribute the observed color shift to the lens.  
Second, the fraction of color-shifted events would be higher in 
the space-based survey because it will monitor much fainter stars 
than current surveys.  As the monitored star becomes fainter, 
the relative flux of the lens to the source star increases and 
thus the event is more likely to produce large color shift.  
Third, with high photometric precision, the survey can detect 
much smaller color shift than ground-based surveys.  With the 
increased fraction of large color-shifted events and higher 
sensitivity to small color shift combined with nearly an order 
higher event rate, the future space-based lensing survey would 
be able to detect color shifts for a large number of events.

In this paper, we demonstrate the usefulness of future space-based 
lensing surveys in measuring color shifts of lensing events.  
For this, we estimate the fraction of events with measurable 
color shifts among the events to be detected from the proposed 
{\it MPF} survey.  We also investigate the properties of the 
color-shifted events by conducting detailed simulation of 
galactic lensing events under realistic observational conditions.  
The fraction of color-shifted events was previously estimated 
by \citet{buchalter96}.  However, their work was intended as 
a template for general comparison with various observational 
assumptions and thus is not focused on a specific survey.  
In addition, they did not provide information about the properties 
of color-shifted events.

The paper is organized as follows.  In \S\ 2, we introduce 
formalism used for the description of the color shift in general 
cases and under various limiting cases.  In \S\ 3, we describe 
the simulation conducted to estimate the fraction of color-shift 
events and their properties.  In \S\ 4, we present the resulting 
distribution of color shift.  We probe the properties of the 
color-shifted events by investigating the distributions of various 
quantities related to the events including the source and lens 
brightness, lens light fraction, magnification, lens mass, and 
event time scale.  In \S\ 5, we summarize and conclude.

\section{Formalism}

The difference between the colors of a lensing event measured 
during and before lensing magnification is expressed as
\begin{equation}
\Delta(R-H)=-2.5 \log \left( {F_R/F_H\over F_{R,0}/F_{H,0}}
\right),
\label{eq1}
\end{equation}
where $F$ represents the observed flux, the subscript `0' 
denotes the measured quantity without lensing magnification, 
and the other subscripts $k=R$ and $H$ denote the observed 
passbands.\footnote{We choose $R$ and $H$ passbands because 
the {\it MPF} survey plans to observe in these bands.  See 
details in \S\ 4.}  The ratio between the observed fluxes 
measured in the individual passbands are 
\begin{equation}
{F_R\over F_H}=
{AF_{R,{\rm S}}+F_{R,{\rm L}}\over AF_{H,{\rm S}}+F_{H,{\rm L}} },
\qquad {F_{R,0}\over F_{H,0}}=
{F_{R,{\rm S}}+F_{R,{\rm L}}\over F_{H,{\rm S}}+F_{H,{\rm L}} },
\label{eq2}
\end{equation}
where $F_{k,{\rm S}}$ and $F_{k,{\rm L}}$ represent the fluxes 
from the source star and lens, respectively.  With some algebra, 
the color shift is represented as 
\begin{equation}
\Delta(R-H)=-2.5 \log \left[
{(A+\eta_R)/(A+\eta_H)\over (1+\eta_R)/(1+\eta_H) } \right],
\label{eq3}
\end{equation}
where $\eta_k$ denotes the ratio between the fluxes from the 
lens and source, i.e.
\begin{equation}
\eta_R={F_{R,{\rm L}}\over F_{R,{\rm S}}},\qquad
\eta_H={F_{H,{\rm L}}\over F_{H,{\rm S}}}.
\label{eq4}
\end{equation}
The sign of the color shift is such that it becomes negative 
when the apparent source star becomes bluer during lensing 
magnification and vice versa.

In the limiting case of a very low magnification ($A\rightarrow 1.0$), 
the amount of the color shift is approximated as
\begin{equation}
\Delta(R-H)\sim -2.5\log\left[1-\epsilon{\eta_R-\eta_H\over
(1+\eta_R)(1+\eta_H)}\right], 
\label{eq5}
\end{equation}
where $\epsilon = A-1.0$.  As $A\rightarrow 1.0$, $\epsilon
\rightarrow 0$, and thus $\Delta(R-H)\rightarrow 0$, implying 
that the color shift expected from a low-magnification event 
is small.

In the opposite limiting case of a very high-magnification 
event ($A\gg 1.0$), the numerator on the right side of 
equation~(\ref{eq3}) become $(A+\eta_R)/(A+\eta_H)\rightarrow 1$, 
and the color shift is approximated as
\begin{equation}
\Delta(R-H)\sim 2.5 \log \left( {1+\eta_R\over 1+\eta_H} \right).
\label{eq6}
\end{equation}
If the lens is very faint (i.e., $\eta_k\ll 1.0$), 
equation~(\ref{eq6}) is further simplified as 
\begin{equation}
\Delta(R-H)\sim  2.5 \log [1+(\eta_R-\eta_H)].
\label{eq7}
\end{equation}
Considering that the lens/source flux ratio is small, $\eta_R-
\eta_H \ll 1.0$ and thus the expected color shift is also 
small.  This implies that high magnification is not a sufficient 
condition to produce noticeable color shift.  To produce noticeable 
color shift, it is additionally required that the lens should 
be bright.  In the limiting case where the lens dominates over 
the source star in brightness (i.e., $\eta_k\gg 1.0$), the 
color shift is represented as
\begin{equation}
\Delta(R-H)\sim 2.5 \log \left( {\eta_R\over \eta_H}\right) 
\equiv (R-H)_{\rm S} - (R-H)_{\rm L},
\label{eq8}
\end{equation}
where $(R-H)_{\rm L}$ and $(R-H)_{\rm S}$ represent the colors
of the lens and source, respectively.  This implies that the 
color shift corresponds to the difference between the colors 
of the lens and source.  Under this condition, the color shift 
is maximized.

\section{Simulation}

We estimate the color-shift distribution of events expected 
from future space-based lensing surveys by conduction Bayesian 
simulation of galactic microlensing events under the observational 
conditions of the {\it MPF} survey.  For this simulation, it is 
required to model of the physical parameters including the mass 
density and velocity distributions of matter along the line of 
sight toward the field, lens mass function, and luminosity 
function of source stars and lenses.  We conduct the simulation 
according to the following procedure.

The survey is assumed to be conducted toward the galactic bulge 
field and the locations of the source and lens along the line 
of sight are allocated based on the spatial mass density distribution.  
Following the specification of the {\it MPF} survey, we set 
the center of the survey field at $(l,b)=(1^\circ\hskip-2pt.2,
-2^\circ\hskip-2pt.4)$.  The spatial mass density distribution 
is modeled by adopting that of \citet{han03}.  In the model, 
the density distribution of bulge matter is based on the deprojected 
infrared light density profile of \citet{dwek95}, specifically model 
G2 with $R_{\rm max}=5$ kpc from their Table 2.  The disk density
distribution is modeled by a double-exponential disk, i.e.
\begin{equation}
\rho(R,z) =\rho_0 
\exp \left[-\left( {R-R_0\over h_R}+{|z|\over h_z}\right)\right], 
\label{eq9}
\end{equation}
where $(R,z)$ are the galactocentric cylindrical coordinates, 
$R_0=8$ kpc is the distance of the Sun from the galactic center, 
$\rho_0= 0.06\ M_\odot \ {\rm pc}^{-3}$ is the local mass density,
and $h_R=3.5$ kpc and $h_z=325$ pc are the radial and vertical 
scale heights, respectively.

Once the locations of the lens and source star are set, we then 
assign their velocities based on the velocity distribution model.
The model velocity distribution has a gaussian form.  The mean 
and dispersion of the bulge velocity distribution are deduced 
from the tensor virial theorem.  The disk velocity distribution 
is assumed to have a flat rotation speed of $v_{\rm c} = 220\ 
{\rm km}\ {\rm s}^{-1}$ and velocity dispersions along and 
normal to the disk plane of $\sigma_\parallel=30\ {\rm km} \ 
{\rm s}^{-1}$ and $\sigma_\perp = 20\ {\rm km}\ {\rm s}^{-1}$, 
respectively.  The resulting distribution of the lens-source 
transverse velocity from the combinations of the bulge and disk 
velocities is listed in Table 1 of \citet{han95}, specifically 
the non-rotating barred bulge model.

We assign the source star brightness by using the luminosity 
function of \citet{holtzman98}.  The Holtzman luminosity 
function is expressed in terms of $V$ band.  Once the $V$-band 
absolute magnitude of each star is chosen from the luminosity 
function, we determine the $R$ and $H$ band magnitude by using 
the color information listed in \citet{cox99}.  The apparent 
magnitudes are then determined considering the distance to 
the source star and extinction.  The extinction is determined 
such that the source star flux decreases exponentially with 
the increase of the dust column density.  The dust column 
density is computed on the basis of an exponential dust 
distribution model with a scale height of $h_z=100\ {\rm pc}$, 
i.e.\ $\propto \exp(-|z|/h_z)$ \citep{drimmel01}.  We set the 
normalization of the extinction in the individual passbands 
such that $A_R=1.8$ and $A_H=0.8$ for a star located at a 
distance $D_{\rm S}=8$ kpc toward the target field.

The mass of the lens is assigned based on the mass function 
of \cite{gould00}.  The model mass function is composed of not 
only stars but also brown dwarfs (BDs) and stellar remnants of 
white dwarfs (WDs), neutron stars (NSs), and black holes (BHs).  
The model is constructed under the assumption that bulge stars 
formed initially according to a double power-law distribution 
of the form
\begin{equation}
{dN\over dM} = k \left( {M\over 0.7\ M_\odot}\right)^\gamma,
\label{eq10}
\end{equation}
where $\gamma=-2.0$ for $M\geq 0.7\ M_\odot$ and $\gamma=-1.3$ 
for $M<0.7\ M_\odot$.  These slopes are consistent with 
\citet{zoccali00} except that the profile is extended below 
their lower limit of $0.15\ M_\odot$ down to a BD cutoff of 
$M=0.03\ M_\odot$.  Based on this initial mass function, 
remnants are modeled by assuming that the stars with initial 
masses $1\ M_\odot< M <8\ M_\odot$, $8\ M_\odot < M < 40\ 
M_\odot$, and $M > 40\ M_\odot$ have evolved into WDs (with 
a mean mass $\langle M\rangle\sim 0.6\ M_\odot$), NSs (with 
$\langle M\rangle \sim 1.35 \ M_\odot$), and BHs (with $\langle 
M\rangle\sim 5\ M_\odot$), respectively.  Then, the resulting 
mass fractions of the individual lens populations are 
${\rm stars}:{\rm BD}:{\rm WD}:{\rm NS}:{\rm BH}=62:7:22:6:3$.  
Once the mass of the lens is set, its brightnesses in $R$ and 
$H$ bands are determined by using the mass-$M_V$ relation and 
the color information presented in \citet{cox99}.  We note 
that BD and remnant lenses are dark and thus events caused by 
them do not produce color shift.

With all the assigned physical parameters of the lens and 
source, we then produce lensing events.  We do this under 
the assumption that the impact parameters of the 
lens-source encounter are randomly distributed.  Since the 
lens flux does not participate in lensing magnification, 
it lowers the apparent magnification of the event, i.e.
\begin{equation}
A_{\rm app}={AF_{\rm S}+F_{\rm L}\over F_{\rm S}+F_{\rm L}};
\qquad
A={u_0+2\over u_0(u_0+4)^{1/2}},
\label{eq11}
\end{equation}
where $u_0$ is the impact parameter of the lens-source 
encounter normalized by the Einstein radius of the lens.
We define detectable events as those with apparent 
magnifications higher than a threshold value of $A_{\rm th}
=3/5^{1/2}$, that corresponds to the magnification when the 
source star just enters the Einstein ring of the lens, i.e. 
$u_0=1.0$.  The apparent magnification varies depending on 
the observed passband.   We use the apparent magnification 
measured in $J$ band as the criterion for event detectability 
because it is the main passband of the {\it MPF} observation
(see below).  We assume that there is no blended flux other 
than the flux from the lens.  We restrict that the survey 
monitors stars with apparent magnitudes brighter than $J=22.5$, 
which corresponds to early M-type main-sequence stars.

Once events are produced, the rate of the individual events 
are weighted by the factor
\begin{equation}
\Gamma \propto \rho(D_{\rm S})D_{\rm S}^2 \rho(D_{\rm L})
r_{\rm E} v,
\label{eq12}
\end{equation}
where $\rho(D)$ is the matter density, $D_{\rm L}$ and 
$D_{\rm S}$ are the distances to the lens and source, 
respectively, the factor $D_{\rm S}^2$ is included to 
account for the increase of the number of source stars with 
the increase of distance, $r_{\rm E}$ is the Einstein radius 
which corresponds to the cross-section of the lens-source 
encounter, and $v$ is the lens-source transverse speed.
The Einstein radius is related to the physical parameters 
of the lens and source by
\begin{equation}
r_{\rm E}=\left[ {4GM\over c^2}
{D_{\rm L}(D_{\rm S}-D_{\rm L})\over D_{\rm S}}\right]^{1/2}.
\label{eq13}
\end{equation}
From the simulation, we find that 66\% and 34\% of the total
events are caused by the lenses located in the bulge and disk, 
respectively.  We also find that events caused by stellar 
lenses comprise 48\% and others are produced by dark lens 
components.

The {\it MPF} survey plans to observe in the wavelength range 
$\sim 600$ -- 1700 nm in three passbands.  These consist of 
a very wide passband spanning the full 600 -- 1700 nm range 
and the other two narrow bands at each end of the range 
\citep{bennett04}.  The very wide filter has no equivalence 
in standard filter, but we assume it corresponds to $J$ filter 
because the central wavelength matches that of the $J$ filter.  
The two narrow filters are approximated by the $R$ and $H$ 
filters. The observing strategy for {\it MPF} is to do most 
of the observations with the very wide filter and a small 
number of observations in the $R$ and $H$ filters. We choose 
color is measured on the images taken in $R$ and $H$ filters 
because larger color shift is expected as the wavelength gap 
between the filters becomes wider.  We assume that {\it MPF} 
can detect 14.3 and 5.0 photons ${\rm s}^{-1}$ for a 22 magnitude 
star in the $R$ and $H$ bands, respectively \citep{bessel79, 
campins85}.

For the individual events, we calculate the maximum color shift 
by using equation~(\ref{eq3}).  The {\it MPF} plans to take 
images with 100 sec exposure at a 2 min interval and combine 
the images to make a 10 min exposure image, resulting in a 
data acquisition rate of $\sim 3$--5 images ${\rm hr}^{-1}$ or  
$\sim 100$ images ${\rm day}^{-1}$.  If 20\% of the observations 
are done in the narrow filters, each of the $R$ and $H$-band 
image is obtained every 2.4 hrs.  This rate is frequent enough 
to cover the peak of nearly all lensing events during which 
the color shift is maximized.  We assume that color shift is 
detectable only if the color shift measured at the peak of 
the event is five times greater than the uncertainty in the 
color-shift measurement, i.e.\ 5$\sigma$ detection.

We determine the uncertainty of the color shift measurement 
as follows.  Let us define ${\cal R}_F$ as the ratio between 
the flux ratios of an event measured in two passbands during 
and before lensing, i.e.,
\begin{equation}
{\cal R}_F={F_R/F_H\over F_{R,0}/F_{H,0}}.
\label{eq14}
\end{equation}
According to the standard error analysis, the fractional 
uncertainty in the measurement of ${\cal R}$ is estimated as
\begin{equation}
{\sigma_{{\cal R}_F}\over {\cal R}_F}  \simeq
\left[
\left({\sigma_{F_{R}/F_{H}}\over F_{R}/F_{H}}\right)^2 +
\left({\sigma_{F_{R,0}/F_{H,0}}\over F_{R,0}/F_{H,0}}\right)^2 
\right]^{1/2},
\label{eq15}
\end{equation}
where the individual terms on the right side are
\begin{equation}
\left( {\sigma_{F_{R}/F_{H}}\over F_{R}/F_{H} } \right)^2
\simeq
\left( {\sigma_{F_R} \over F_R}\right)^2 + 
\left({\sigma_{F_H} \over F_H} 
\right)^2,
\label{eq16}
\end{equation}
and 
\begin{equation}
\left( {\sigma_{F_{R,0}/F_{H,0}}\over F_{R,0}/F_{H,0} } \right)^2 
\simeq
\left( {\sigma_{F_{R,0}} \over F_{R,0}}\right)^2 + 
\left({\sigma_{F_{H,0}} \over F_{H,0}} 
\right)^2.
\label{eq17}
\end{equation}
Under the assumption that the photometry is photon dominated, 
i.e. $\sigma_F=F^{1/2}$, equations~(\ref{eq16}) and (\ref{eq17}) 
are approximated as 
\begin{equation}
\left( {\sigma_{F_{R}/F_{H}}\over F_{R}/F_{H} }\right)^2 \sim
{1\over F_R}+{1\over F_H} =
{F_R+F_H \over F_R F_h},
\label{eq18}
\end{equation}
and 
\begin{equation}
\left( {\sigma_{F_{R,0}/F_{h,0}}\over F_{R,0}/F_{H,0} } \right)^2 
\sim {1\over F_{R,0}}+{1\over F_{H,0}} =
{F_{R,0}+F_{H,0} \over F_{R,0} F_{H,0}}.
\label{eq19}
\end{equation}
Then, from equation~(\ref{eq15}), (\ref{eq18}), and (\ref{eq19}), 
one finds 
\begin{equation}
{\sigma_{{\cal R}_F}\over {\cal R}_F}  \simeq
\left( {F_R+F_H \over F_R F_H} +
{F_{R,0}+F_{H,0} \over F_{R,0} F_{H,0}}  \right)^{1/2}.
\label{eq20}
\end{equation}
The uncertainty in units of magnitude is related to the 
fractional uncertainty $\sigma_{{\cal R}_F}/ 
{\cal R}_F$ by
\begin{equation}
\sigma_{\Delta(R-H)}=2.5 \log e 
\left( { \sigma_{{\cal R}_F}\over {\cal R}_F} \right)
\sim 1.1 \left( { \sigma_{{\cal R}_F}\over {\cal R}_F} \right).
\label{eq21}
\end{equation}
Then, the color-shift uncertainty is expressed as
\begin{equation}
\sigma_{\Delta(R-H)} \sim 1.1
\left( {F_R+F_H \over F_R F_H} +
{F_{R,0}+F_{H,0} \over F_{R,0} F_{H,0}}  \right)^{1/2}.
\label{eq22}
\end{equation}

\begin{figure}[t]
\epsscale{1.18}
\plotone{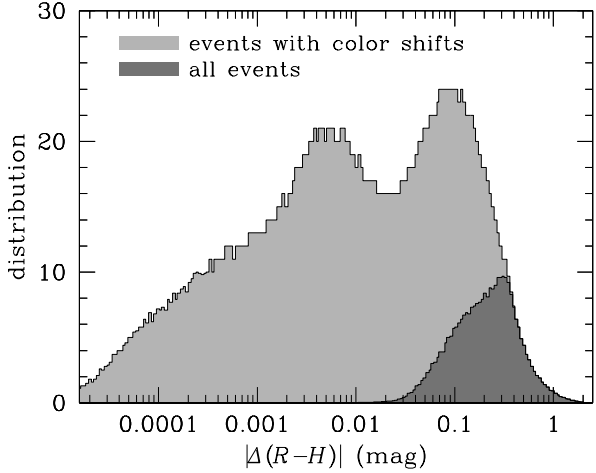}
\caption{\label{fig:one}
Color shift distribution of events expected to be detected from
the future {\it MPF} lensing survey.  Darker shade region represents 
the distribution for events where the color shift can be detected 
at 5$\sigma$ level.  
}\end{figure}

\section{Result}

In Figure~\ref{fig:one}, we present the color shift distribution 
of events expected to be detected from the {\it MPF} survey.  
From the distribution, we find that color shift can be measured 
for $\sim 12\%$ of the total events.  Considering that events 
produced by stellar lenses comprise $\sim 48\%$ of the total event, 
this corresponds to $\sim 23\%$ of all events produced by stellar 
lenses.  We find that the fractions of color-shifted events among 
the disk and bulge events are $\sim 18\%$ and 9\%, respectively.  
The fraction for the disk-lens events is higher because the lenses 
are located closer to the observer and thus they are brighter.

\begin{figure}[th]
\epsscale{1.18}
\plotone{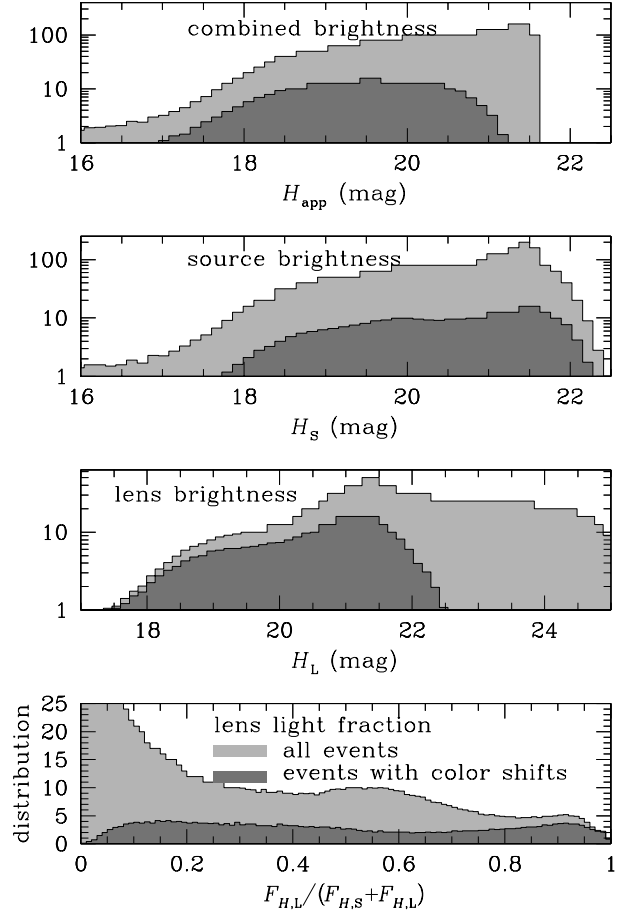}
\caption{\label{fig:two}
Top panel shows the distribution of the apparent brightness 
of the combined image of the source star and lens.  Second 
and third panels are the brightness distributions of the 
separate source stars and lenses, respectively.  Bottom 
panel shows the distribution of the $H$-band lens light 
fraction.  The lens light fraction is 1.0 when the observed 
flux is purely from the lens.  Dark shade regions represent 
the distributions for events where the color shift is detectable 
at 5$\sigma$ level.
}\end{figure}

We investigate the properties of the color-shifted events.
First, we examine the brightness of the source stars and lenses.
In Figure~\ref{fig:two}, we present the $H$-band brightness 
distributions of the source+lens combined images (top panel), 
and separate source stars and lenses (second and third panels, 
respectively).  From the distributions, we find that the 
color-shifted events have apparent brightness distributed in 
the range of $17\lesssim H\lesssim 21$.   We also find that 
the source and lens stars of color-shifted events have similar 
brightness ranges of $18\lesssim H\lesssim 22$.  This implies 
that color shift occurs when the lens brightness is equivalent 
to that of the source star.  This can also be seen in the bottom 
panel of Figure~\ref{fig:two}, where we present the distribution 
of the $H$-band lens light fraction.  We find that $\sim 45\%$ 
of the color-shifted events will be produced by lenses brighter 
than source stars in the $H$ band and $\sim 11\%$ of them will 
have lens light fractions greater than $F_{H,{\rm L}}/(F_{H,{\rm S}} 
+ F_{H,{\rm L}})=0.9$.

\begin{figure}[t]
\epsscale{1.18}
\plotone{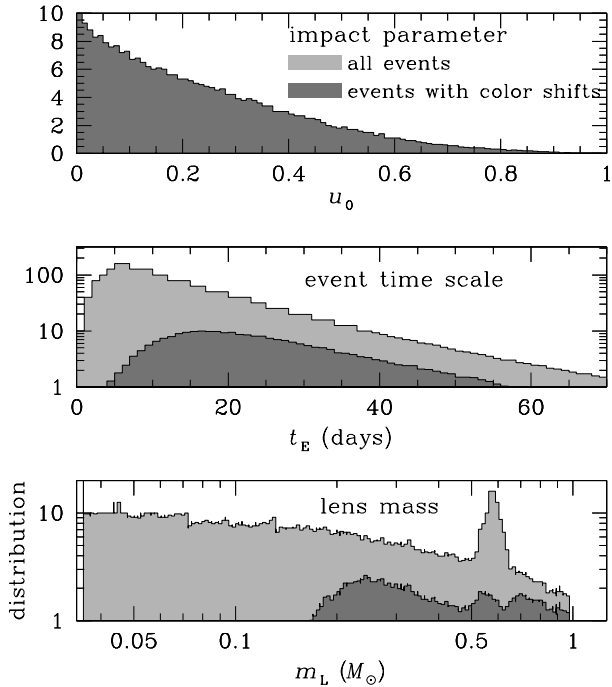}
\caption{\label{fig:three}
Distributions of the normalized impact parameters of the 
lens-source encounter (top), event time-scale (middle), and 
lens masses (bottom) of events to be detected by the future 
{\it MPF} lensing survey.  Dark shade region represents 
the distribution for events where the color shift is detectable 
at 5$\sigma$ level.
}\end{figure}

Second, we investigate the magnifications of the color-shifted
events.  For this, we construct the distribution of the impact 
parameters of the lens-source encounter.  The distribution is 
presented in the top panel of Figure~\ref{fig:three}.  From 
the distribution, we find, as expected, that color-shifted 
events tend to have high magnifications.  We estimate that the 
fraction of the color-shifted events with impact parameters 
less than $u_0=0.1$ (or $A\gtrsim 10$) is $\sim 31\%$.  On the 
other hand, the fraction of the color-shifted events among 
those with $u_0> 0.5$ (or $A<2.4$) is merely $\sim 10\%$.  
These results imply that color shifts will be detected mainly 
from high-magnification events.

Third, we investigate the event time scales.  The event time 
scale is defined as the time required for the source star to 
cross the Einstein radius of the lens, i.e.\ $t_{\rm E}=r_{\rm E}/v$.  
The middle panel of Figure~\ref{fig:three} shows the distribution 
of the time scale.  By comparing the distribution with that of 
all events, we find that the time scales of the color-shifted 
events tend to be longer than the events without color shifts.  
The most frequent value of the time scale for the color-shifted 
events is $\sim 18$ days.  On the other hand, the mode value of 
all events is $\sim 6$ days.  We find this tendency is because 
the lenses of the color-shifted events tend to be brighter and 
thus heavier than the lenses of events without color shifts.  
We find that the lenses responsible for the color-shifted events 
have masses located in the range of $0.2\ M_\odot\lesssim M 
\lesssim 1.0\ M_\odot$.  These correspond to stars with spectral 
types from G down to mid M-type main-sequence stars.  Color shift 
measurement for events produced by late spectral-type lens stars 
is possible because the {\it MPF} survey will observe in near 
infrared through which the lens/source flux ratio is not negligible.

The selection effect of color-shift measurements that favors 
events with higher magnifications and longer time scales 
implies that the fraction of planetary microlensing events 
with color-shift measurements would be substantially higher 
than the average value of all events. This is because planet 
detection has a similar selection effect of preferring events 
with higher magnifications and longer time scales \citep{han07}.  
Therefore, it would be possible to characterize the host stars 
of planets from color-shift measurements for an important 
fraction of planetary events. For some cases, the source and 
lens stars may have similar colors resulting in small color 
shifts. In this case, the lens stars can still be detectable 
with space-based follow-up observations conducted several years 
after the event \citep{bennett07}.

\section{Conclusion}

We investigated the usefulness of future space-based lensing 
surveys in measuring color shifts of microlening events caused 
by the flux from the lens.  By conducting simulation of galactic 
microlensing events based on the specification of the proposed 
{\it MPF} survey, we estimated that the shift in the color of 
$R-H$ would be measured at 5$\sigma$ level for $\sim 12\%$ of 
events that would occur on source stars with apparent magnitudes
brighter than $J=22.5$.  Color-shifted events tend to have high 
magnifications and the lenses would have brightnesses equivalent 
to those of source stars. We estimated that $\sim 31\%$ of the 
color-shifted events would have magnifications $A > 10$ and the 
lenses of $\sim 45\%$ of the color-shifted events would be 
brighter than source stars.  The time scales of the color-shifted 
events would be longer than those without color shifts because 
these events tend to be brighter and thus heavier than those 
without color shifts.  We estimated that the most frequent time 
scale of the color-shifted events would be $\sim 18$ days, while 
the mode value of all events would be $\sim 6$ days.  The spectral 
type of the lens stars of the color-shifted events would extend 
down to mid M type because the {\it MPF} survey will observe in 
near infrared through which the lens/source flux ratio is not 
negligible.  Being able to extract information about the lens 
from the measured color shift, future space-based lensing surveys 
would be able to better constrain the lens population of galactic 
microlensing events.

\acknowledgments 
This work was supported by the Basic Research Fund of Chungbuk 
National University.


\begin{thebibliography}{99}
\frenchspacing
\bibitem[Bennett et al.(2004)]{bennett04}
Bennett, D.\ P., et al.\ 2004, Proc. SPIE Int. Soc. Opt. Eng., 5487, 1453

\bibitem[Bennett, Anderson \& Gaudi]{bennett07}
Bennett, D.\ P., Anderson, J., \& Gaudi, B. S.\ 2007, \apj, 660, 781

\bibitem[Bennett \& Rhie(2002)]{bennett02}
Bennett, D.\ P., \& Rhie, S.\ H.\ 2002, \apj, 574, 985

\bibitem[Bessel(1979)]{bessel79}
Bessel, M.\ S. 1979, \pasp, 91, 589

\bibitem[Buchalter et al.(1996)]{buchalter96}
Buchalter, A., Kamionkowski, M., \& Rich, R.\ M.\ 1996, \apj, 469, 676

\bibitem[Campins et al.(1985)]{campins85}
Campins, H., Reike, G.\ H., \& Lebofsky, M.\ J.\ 1985, \aj, 90, 896

\bibitem[Cox(1999)]{cox99}
Cox, A.\ N., 1999, Allen's Astrophysical Quantities 
(4th Ed.; New York: Springer)

\bibitem[Drimmel \& Spergel(2001)]{drimmel01}
Drimmel, R., \& Spergel, D.\ N.\ 2001, \apj, 556, 181

\bibitem[Dwek et al.(1995)]{dwek95}
Dwek, E., et al.\ 1995, \aj, 445, 716

\bibitem[Gould(2000)]{gould00}
Gould, A.\ 2000, \apj, 535, 928

\bibitem[Han(2007)]{han07}
Han, C.\ 2007, \apj, 661, 1202

\bibitem[Han \& Gould(1995)]{han95}
Han, C., \& Gould, A.\ 1995, \apj, 447, 53

\bibitem[Han \& Gould(2003)]{han03}
Han, C., \& Gould, A.\ 2003, \apj, 592, 172

\bibitem[Holtzman et al.(1998)]{holtzman98}  
Holtzman, J.\ A., Watson, A.\ M., Baum, W.\ A., Grillmair, C.\ J., 
Groth, E.\ J., Light, R.\ M., Lynds, R., \& O'Neil, E.\ J.\ 1998, \aj, 
115, 1946

\bibitem[Kamionkowski(1995)]{kamionkowski95}
Kamionkowski, M.\ 1995, \apj, 442, L9

\bibitem[Sumi(2004)]{sumi04}
Sumi, T.\ 2004, \mnras, 349, 193

\bibitem[Zoccalli et al.(2000)]{zoccali00} 
Zoccali, M.\ S., Cassisi, S., Frogel, J.\ A., Gould, A., Ortolani, S., 
Renzini, A.,  Rich, R.\ M. 1999, \& Stephens, A.\ 2000, \apj, 530, 418

\end{thebibliography}
\end{document}